\documentclass[conference]{IEEEtran}

\usepackage{amsmath,amssymb}
\usepackage{graphicx}
\usepackage{booktabs}
\usepackage{url}
\usepackage{hyperref}
\usepackage{cite}
\usepackage{subcaption}
\usepackage{stfloats}   
\usepackage{tikz}
\usepackage{listings}
\usepackage{xcolor}

\usetikzlibrary{positioning}

\hypersetup{
    colorlinks=true,
    linkcolor=black,
    citecolor=black,
    urlcolor=blue
}

\begin{document}

\title{SecureCodeRL: Security-Aware Reinforcement Learning for Code Generation with Partial-Credit Rewards}

\author{
\IEEEauthorblockN{Suryansh Singh Sijwali}
\IEEEauthorblockA{\textit{Department of Computer Science and Engineering} \\
\textit{The Pennsylvania State University}\\
University Park, PA 16802, USA \\
sss6371@psu.edu}
\and
\IEEEauthorblockN{Suman Saha}
\IEEEauthorblockA{\textit{Department of Computer Science and Engineering} \\
\textit{The Pennsylvania State University}\\
University Park, PA 16802, USA \\
sumsaha@psu.edu}
}

\maketitle

\begin{abstract}
Large Language Models (LLMs) can generate plausible code, but in settings that require exact stdin/stdout behavior they frequently produce programs that compile yet fail tests, and in some cases they introduce security-sensitive patterns. This paper presents SecureCodeRL, a reinforcement learning (RL) pipeline for security-aware code generation that optimizes a combined reward $R=\alpha R_{\text{func}}+\beta R_{\text{sec}}$. The key idea is a \emph{partial-credit} functional reward that assigns intermediate scores for syntactic validity, successful execution, and producing output, reducing reward sparsity that otherwise stalls learning on competitive-programming style tasks. I evaluate supervised fine-tuning (SFT) and PPO variants on a small held-out prompt set from APPS+ and observe that PPO with partial credit (using a continued-training variant) improves syntax validity from 45\% (SFT) to 60\% and achieves the only non-zero test success signal in this pilot evaluation (5\% at-least-one-test-pass), while remaining 100\% clean under Bandit static analysis. Although Bandit findings were absent in this small evaluation, the security term is integrated into training to discourage insecure shortcuts when they appear.
\end{abstract}


\section{Introduction}

Code-capable LLMs are now widely used as programming assistants, but ``looks correct'' is not the same as \emph{is correct}. In competitive-programming style problems, the program must parse input exactly, implement the right algorithm, and print output in the expected format. In practice, many generations fail due to wrong output, timeouts, crashes, or even producing no output at all. In parallel, there is a security angle: prior work has found that model suggestions can include insecure patterns and risky APIs \cite{pearce2022asleep}.

This project grew out of a simple question I kept running into while benchmarking code models: \emph{If the reward is ``all tests pass,'' how does an RL system learn anything when almost nothing passes?} In my APPS+ runs, the most common failures were not dramatic crashes; they were ``nearly-right'' programs that were off by formatting, missed a print statement, or computed a value but never emitted it. A binary pass/fail reward treats these failures the same as syntax errors, which makes reward sparse and learning brittle.

SecureCodeRL is my attempt to make that learning signal less fragile. This approach optimizes two objectives simultaneously:
\begin{itemize}
    \item I optimize for \textbf{functional correctness} using test execution feedback.
    \item I incorporate \textbf{security awareness} using static analysis (Bandit) as a penalty term.
\end{itemize}
To bridge the gap between ``fails everything'' and ``passes everything,'' I introduce a \textbf{partial-credit} $R_{\text{func}}$ that explicitly rewards intermediate milestones that matter in stdin/stdout judges.

The full codebase for this project is publicly available.\footnote{Repository: \url{https://github.com/SuryanshSS1011/basic-rl-feedback-workflow}}

\subsection{Contributions}

I make the following contributions:

\begin{itemize}
    \item \textbf{Evidence-driven motivation via benchmarking.} I evaluate multiple open code models on APPS+ and quantify how often they produce syntactically valid code, pass tests, and remain security-clean.
    \item \textbf{A joint objective for correctness and security.} I train with $R=\alpha R_{\text{func}}+\beta R_{\text{sec}}$, where $R_{\text{sec}}$ is derived from Bandit findings.
    \item \textbf{Partial-credit reward shaping.} I design an execution-stage reward that distinguishes syntax errors from runnable-but-silent programs (the ``missing print'' pattern) and from partial test matches.
    \item \textbf{Pilot RL results.} On a small held-out prompt set, PPO with partial credit improves syntax validity and is the only evaluated variant to achieve non-zero test success while remaining Bandit-clean.
\end{itemize}

\section{Background and Related Work}

\subsection{Code generation and execution feedback}

Recent code LLMs (e.g., DeepSeek-Coder, Code Llama, StarCoder2) are trained primarily with next-token prediction on large code corpora and can produce fluent solutions, but correctness under strict unit tests remains challenging \cite{deepseekcoder2024,codellama2023,starcoder2_2024}. A natural direction is to close the loop with compiler/test feedback. Prior work has explored reinforcement learning for code generation using execution signals, including CodeRL \cite{coderl2022} and reinforcement learning from unit test feedback \cite{rltf2023}. These ideas motivate SecureCodeRL: define reward using tools that directly reflect what a judge or developer cares about.

\subsection{Security analysis as a training signal}

Static analyzers provide an operational way to discourage insecure patterns. In Python, Bandit flags a range of risky constructs (e.g., use of \texttt{eval}, unsafe subprocess usage, weak cryptography) \cite{bandit}. In SecureCodeRL, I treat static-analysis findings as a penalty shaping term: code that passes tests but triggers serious findings should receive a lower reward.

\section{Dataset, Benchmark, and Failure Modes}

\subsection{Dataset}

I use the APPS family of competitive programming problems \cite{apps2021}, and specifically the APPS+ distribution used in my implementation \cite{appsplus_repo} (the \texttt{APPS\_Plus} release).\footnote{Direct data file used in my pipeline: \url{https://raw.githubusercontent.com/Ablustrund/APPS_Plus/refs/heads/main/data/v1/data.json}.} Each problem includes a natural language statement and multiple stdin/stdout test cases. This stdin-style format is exactly where small I/O mistakes (missing prints, extra labels, whitespace) can cause total failure.

\subsection{Multi-model benchmark: the gap between ``valid'' and ``correct''}
\label{sec:benchmark}

Before training SecureCodeRL, I benchmarked several models to quantify baseline difficulty. Table~\ref{tab:benchmark} reports the summary table I used to motivate this project: even when models generate syntactically valid code, \emph{full} test success remains low. This gap between \emph{parses} and \emph{passes} is the core reason I wanted an RL loop driven by execution. (Later, in my pilot RL evaluation, I also report an ``at least one test passes'' indicator to detect the first signs of functional learning at small $N$.)

\begin{table}[t]
\centering
\caption{Benchmark summary on stdin-style APPS+ subset (my runs). \emph{Test Pass} denotes the percentage of prompts where the generated program passes \textbf{all} provided tests. \emph{Security Clean} is computed using Bandit findings in my pipeline (with an additional lightweight regex heuristic used during benchmarking for debugging); Bandit is the primary metric reported.}
\label{tab:benchmark}
\begin{tabular}{lccc}
\toprule
Model & Syntax Valid & Test Pass & Security Clean \\
\midrule
DeepSeek-6.7B & 83.4\% & 14.3\% & 96.3\% \\
CodeLlama-7B  & 48.9\% & 7.6\%  & 99.5\% \\
StarCoder2-7B & 20.2\% & 1.3\%  & 99.7\% \\
\bottomrule
\end{tabular}
\end{table}

The headline is not that ``models are bad''---it is that the learning signal is \emph{sparse}. If only a small fraction of generations ever fully pass, then binary reward ($1$ if all tests pass, else $0$) provides almost no gradient. That is exactly the regime where reward shaping matters.

\subsection{Failure taxonomy: what actually goes wrong}
\label{sec:taxonomy}

To design a reward that helps PPO learn in this regime, I looked at \emph{how} generations fail under the test harness. The dominant outcomes fall into a small set of categories shown in Table~\ref{tab:failtax}. The key point is that many failures are ``near misses'' (especially wrong output and no output), which a binary reward collapses into the same zero as a syntax error. These proportions come from my benchmark logs over the stdin-style APPS+ subset (Section~\ref{sec:benchmark}).

\begin{table}[t]
\centering
\caption{Compact failure taxonomy observed during my benchmark analysis (approximate shares).}
\label{tab:failtax}
\begin{tabular}{lc}
\toprule
Failure type & Share of failures \\
\midrule
Wrong output   & $\sim$60\% \\
No output      & $\sim$25\% \\
Timeout/hang   & $\sim$8\%  \\
Crash          & $\sim$5\%  \\
Memory error   & $\sim$2\%  \\
\bottomrule
\end{tabular}
\end{table}

This taxonomy is the bridge from benchmark to reward design: partial credit is explicitly shaped to separate \emph{syntax}, \emph{runtime stability}, and \emph{I/O behavior} (the no-output bucket) before demanding full correctness.

\subsection{Qualitative failure modes (representative examples)}

Numbers are helpful, but for code generation the patterns are often clearer in short snippets.

\subsubsection{Failure mode A: the ``missing print'' bug (silent programs)}

A frequent pattern is code that reads input and computes a value, but never prints it. Under binary reward this is indistinguishable from a syntax error.

\begin{lstlisting}
# Reads input and computes, but produces NO OUTPUT
n = int(input())
(n * 2)   # expression is evaluated and discarded
\end{lstlisting}

\subsubsection{Failure mode B: wrong-format output}

Some generations print output, but with labels or extra text (strict judges reject it):

\begin{lstlisting}
n = int(input())
print("Result:", n * 2)   # wrong format for a judge
\end{lstlisting}

\subsubsection{Failure mode C: insecure shortcuts (e.g., \texttt{eval})}

In a minority of cases, models reach for insecure shortcuts. Even if a shortcut ``works,'' I want the training signal to discourage it.

\begin{lstlisting}
formula = "a + b"
out = eval(formula)  # risky shortcut; Bandit flags this pattern
\end{lstlisting}

\section{Method}

\subsection{Problem formulation}

Given a prompt $x$, a code model $\pi_\theta$ generates a program $y \sim \pi_\theta(\cdot \mid x)$. I define a scalar reward:

\begin{equation}
R(y) = \alpha \cdot R_{\text{func}}(y) + \beta \cdot R_{\text{sec}}(y),
\end{equation}

with $\alpha = 0.6$ and $\beta = 0.4$ in my experiments. I choose these weights to slightly prioritize functional correctness while still imposing a meaningful penalty for insecure patterns. The functional term measures correctness under test execution, and the security term penalizes static-analysis findings.

\subsection{Security reward: $R_{\text{sec}}$}

I run Bandit on generated Python code and map findings to a penalty score in $[0,1]$. In the pilot evaluation reported later, all generations were Bandit-clean, so $R_{\text{sec}}=1.0$ across models. The framework is still useful: when insecure patterns appear (e.g., \texttt{eval}, unsafe subprocess usage), the reward would drop, pushing the policy away from those shortcuts \cite{bandit}.

\subsection{Functional reward: why binary feedback is not enough}

A strict functional reward can be written as:

\[
R_{\text{func}}(y) =
\frac{\#\text{tests passed}}{\#\text{tests total}}.
\]

This is principled, but in the regime suggested by Table~\ref{tab:benchmark} it becomes effectively sparse: many generations fail all tests, and the model cannot easily learn the difference between ``almost correct'' and ``nonsense.''

\subsection{Partial-credit functional reward (key idea)}

The failure taxonomy in Table~\ref{tab:failtax} makes the design goal concrete. If a large chunk of failures are \emph{no output}, I need the reward to explicitly reward output production. If a chunk are \emph{crashes}, I want to reward runtime stability. I therefore define $R_{\text{func}}$ as a staged score in $[0,1]$ that distinguishes:
\begin{itemize}
    \item syntax validity,
    \item running without runtime error,
    \item producing any stdout,
    \item partial test matches,
\end{itemize}
with full credit reserved for passing tests.

\begin{table}[t]
\centering
\caption{Partial-credit functional reward used in our implementation. Let $T$ be the number of tests and $k$ the number of tests passed. If a program produces no stdout, we treat it as failing the output stage even if it runs.}
\label{tab:rfunc}
\begin{tabular}{llc}
\toprule
Stage & Condition & $R_{\text{func}}$ \\
\midrule
0 & Syntax error / not runnable & $0.0$ \\
1 & Parses as Python (valid syntax) & $0.2$ \\
2 & Executes without runtime error & $0.4$ \\
3 & Produces any stdout & $0.6$ \\
4 & Passes tests (partial or full) & $0.6 + 0.4\cdot \frac{k}{T}$ \\
\bottomrule
\end{tabular}
\end{table}


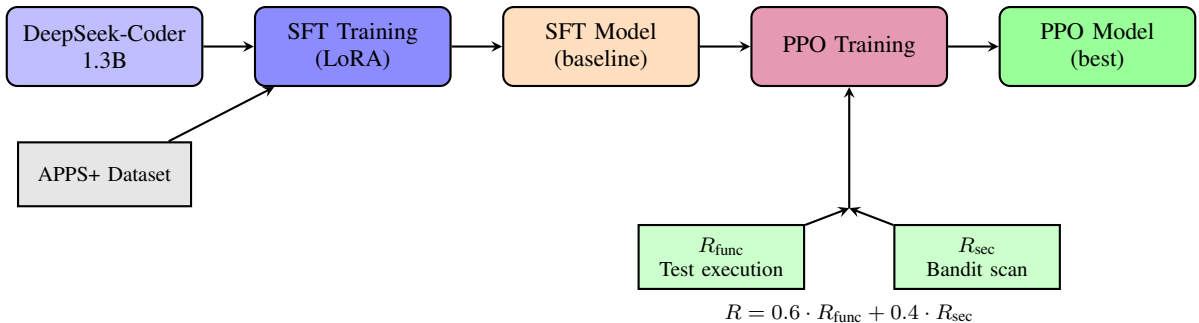
\begin{figure*}[t]
\centering
\begin{tikzpicture}[
    box/.style={rectangle, draw=black, thick, rounded corners, minimum width=2.6cm, minimum height=1.05cm, align=center, font=\small},
    data/.style={rectangle, draw=black, thick, fill=gray!20, minimum width=2.3cm, minimum height=0.85cm, align=center, font=\footnotesize},
    reward/.style={rectangle, draw=black, thick, fill=green!20, minimum width=2.2cm, minimum height=0.85cm, align=center, font=\footnotesize},
    arrow/.style={->, thick, >=stealth}
]
\node[box, fill=blue!25] (base) at (0,0) {DeepSeek-Coder\\1.3B};
\node[data] (apps) at (0,-1.7) {APPS+ Dataset};
\node[box, fill=blue!45] (sft) at (3.3,0) {SFT Training\\(LoRA)};
\node[box, fill=orange!25] (sft_model) at (6.6,0) {SFT Model\\(baseline)};

\node[box, fill=purple!40] (ppo) at (9.9,0) {PPO Training};
\node[box, fill=green!40] (final) at (13.2,0) {PPO Model\\(best)};

\node[reward] (rfunc) at (8.2,-2.8) {$R_{\text{func}}$\\Test execution};
\node[reward] (rsec)  at (11.6,-2.8) {$R_{\text{sec}}$\\Bandit scan};

\coordinate (merge) at (9.9,-2.15);

\draw[arrow] (base) -- (sft);
\draw[arrow] (apps) -- (sft);
\draw[arrow] (sft) -- (sft_model);
\draw[arrow] (sft_model) -- (ppo);
\draw[arrow] (ppo) -- (final);

\draw[arrow] (rfunc) -- (merge);
\draw[arrow] (rsec)  -- (merge);
\draw[arrow] (merge) -- (ppo);

\node[font=\footnotesize] at (9.9,-3.55) {$R = 0.6\cdot R_{\text{func}} + 0.4\cdot R_{\text{sec}}$};
\end{tikzpicture}
\caption{SecureCodeRL pipeline: SFT (LoRA) initializes a policy, then PPO optimizes a combined reward. Functional and security rewards merge into a single signal before updating the PPO policy.}
\label{fig:pipeline}
\end{figure*}

\begin{figure*}[t]
\centering
\begin{tikzpicture}[
    box/.style={rectangle, draw=black, thick, minimum width=2.8cm, minimum height=0.8cm, align=center, font=\small},
    score/.style={circle, draw=black, thick, fill=blue!20, minimum size=0.8cm, font=\small\bfseries},
    arrow/.style={->, thick, >=stealth}
]
\node[box, fill=red!20]    (s0) at (0,0)   {Syntax Error};
\node[box, fill=orange!20] (s1) at (3.5,0) {Valid Syntax};
\node[box, fill=yellow!20] (s2) at (7,0)   {Runs};
\node[box, fill=green!20]  (s3) at (10.5,0){Output};
\node[box, fill=blue!20]   (s4) at (14,0)  {Tests Pass};

\node[score] at (0,-1.2)   {0.0};
\node[score] at (3.5,-1.2) {0.2};
\node[score] at (7,-1.2)   {0.4};
\node[score] at (10.5,-1.2){0.6};
\node[score] at (14,-1.2)  {1.0};

\draw[arrow] (s0) -- (s1);
\draw[arrow] (s1) -- (s2);
\draw[arrow] (s2) -- (s3);
\draw[arrow] (s3) -- (s4);

\node[above of=s0, node distance=1cm, font=\footnotesize\itshape] {Stage 0};
\node[above of=s1, node distance=1cm, font=\footnotesize\itshape] {Stage 1};
\node[above of=s2, node distance=1cm, font=\footnotesize\itshape] {Stage 2};
\node[above of=s3, node distance=1cm, font=\footnotesize\itshape] {Stage 3};
\node[above of=s4, node distance=1cm, font=\footnotesize\itshape] {Stage 4};

\node[font=\footnotesize] at (7,-2.2) {$R_{\text{func}}$ partial-credit stages (staged scoring)};
\end{tikzpicture}
\caption{Partial-credit functional reward: instead of binary test pass/fail, I assign intermediate scores for syntax validity, successful execution, and producing output, with full credit reserved for passing tests.}
\label{fig:partialcredit}
\end{figure*}
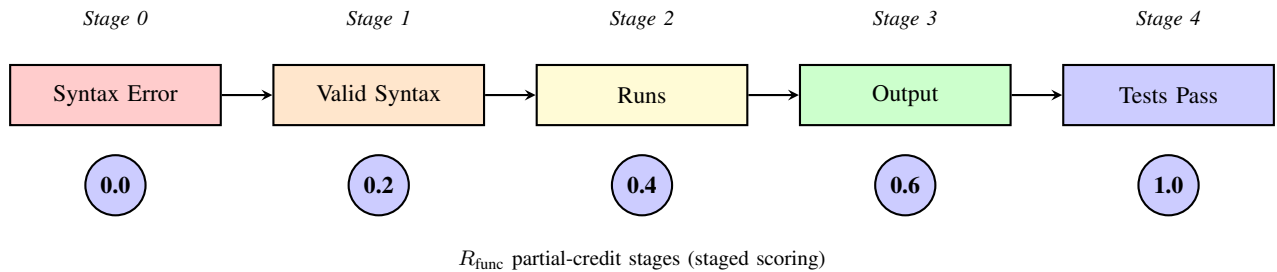

\section{Experiments}

\subsection{Models and training}

I start from DeepSeek-Coder-1.3B-Instruct \cite{deepseekcoder2024} and adapt it with LoRA. I train an SFT baseline on stdin-style APPS+ problems, then run PPO variants initialized from SFT (and, in one variant, continued from a prior PPO checkpoint). In particular, \textbf{PPO-continue} resumes training from a PPO-simple checkpoint to test whether partial-credit shaping accumulates benefits over multiple PPO phases. The overall training flow is summarized in Fig.~\ref{fig:pipeline}.

\subsection{Evaluation protocol and metrics}

For evaluation, I run an execution harness on a held-out set of prompts and compute:
\begin{itemize}
    \item \textbf{Syntax Valid \%}: parses/compiles as Python.
    \item \textbf{At-least-one-test-pass \%}: fraction of prompts where the generated program passes \emph{at least one} test case.
    \item \textbf{Security Clean \%}: no Bandit findings.
    \item \textbf{Mean rewards}: $R_{\text{func}}$, $R_{\text{sec}}$, and combined $R$.
\end{itemize}

I use the ``at least one test passes'' metric in this pilot because it is sensitive to early signs of learning when $N$ is small; in a scaled evaluation, I would also report (and emphasize) the stricter all-tests-pass rate used in the benchmark in Section~\ref{sec:benchmark}.

My evaluation is intentionally small and focused (20 prompts per model) to validate the pipeline end-to-end and to check whether partial credit produces non-trivial learning signals. I treat the resulting numbers as a pilot study rather than a final benchmark.

\section{Results}

\subsection{Quantitative comparison (pilot)}

Table~\ref{tab:results} summarizes my evaluation results (20 prompts per model). The headline result is that PPO with partial credit (continued from PPO-simple) achieves the highest syntax validity (60\%) and is the only evaluated model with a non-zero test success signal (5\% at-least-one-test-pass) under my protocol, while remaining 100\% Bandit-clean.

\begin{table}[t]
\centering
\caption{SecureCodeRL evaluation results (20 prompts per model). Note: unlike Table~\ref{tab:benchmark} (all-tests-pass), this table reports $\ge$1 test pass to capture early learning signals in a small-$N$ pilot.}
\label{tab:results}
\begin{tabular}{lcccc}
\toprule
Model & Syntax \% & $\ge$1 Test Pass \% & Security \% & Mean $R$ \\
\midrule
SFT Baseline        & 45.0 & 0.0 & 100.0 & 0.40 \\
PPO-simple          & 15.0 & 0.0 & 100.0 & 0.40 \\
PPO-fresh           & 25.0 & 0.0 & 100.0 & 0.40 \\
\textbf{PPO-continue} & \textbf{60.0} & \textbf{5.0} & 100.0 & \textbf{0.41} \\
\bottomrule
\end{tabular}
\end{table}

A detail worth emphasizing is that the absolute test success rate is still low. In this evaluation, ``5\%'' corresponds to one prompt for which the model passed at least one test. That said, the fact that \emph{only} the partial-credit PPO-continue variant crosses zero is consistent with the original motivation: binary rewards can keep PPO stuck, and partial credit can help it reach the first rung of functional success.

\subsection{Security results: Bandit vs.\ heuristic findings}

All evaluated generations were Bandit-clean, so $R_{\text{sec}}$ did not differentiate models in this pilot. I also tracked a lightweight regex-based heuristic counter during experiments; those heuristic findings increased in PPO-continue, which I interpret as a byproduct of longer/more varied generations triggering simplistic patterns. For reporting security in the paper, I treat Bandit as the reference and heuristics as debugging tools.

\section{Discussion}

\subsection{Why partial credit helped in this setting}

The partial-credit design follows directly from the failure taxonomy in Table~\ref{tab:failtax}. If a model reads input and computes a value but fails to print it, a strict test reward returns zero, even though the fix is small. Partial credit gives these programs a non-zero score (syntax $\rightarrow$ runs $\rightarrow$ output) and makes it easier for PPO to climb the ladder instead of restarting from scratch each episode.

Another practical benefit is that staged rewards help me debug training. When the reward is always zero, it is hard to tell whether the model is making progress on syntax, runtime stability, or I/O behavior. With partial credit, the reward itself becomes a diagnostic signal.

\subsection{What I still need for stronger validation}
\label{sec:validation_next}

The pilot results in Table~\ref{tab:results} are enough to validate the pipeline end-to-end and to show that partial credit changes PPO's behavior in the direction I intended: PPO-continue is the only variant in my evaluation that achieves non-zero test success, and it substantially improves syntax validity. At the same time, I do not want to over-interpret what a 20-prompt study can establish. To strengthen the empirical validation, I see three concrete upgrades.

First, I need a larger and more systematic evaluation. The most direct change is to scale from $N=20$ prompts per model to at least a few hundred, stratified by APPS+ difficulty. With small $N$, a single success can move the test metric by multiple points, so the current estimate has high variance.

Second, I need uncertainty estimates and controlled ablations. I plan to report bootstrap confidence intervals for Syntax \% and test success, and I will add comparisons that isolate each design choice: PPO with a binary functional reward vs.\ PPO with partial credit; partial credit with and without the security term; and security shaping with a binary functional reward. These ablations directly test whether the improvements come from reward design rather than initialization effects or noise.

Third, I need stronger baselines and broader security coverage. On the functionality side, I should compare against inference-time baselines such as best-of-$k$ sampling with test-based selection, to separate the value of training from simply sampling more candidates. On the security side, Bandit is a good starting point for Python, but adding additional analyzers (or a curated vulnerability suite) would make the security claim more robust.

Finally, I will include a short qualitative appendix with paired examples that make the reward shaping behavior explicit: a missing-print program fixed into a correct one; an output-formatting failure corrected; a timeout avoided; and at least one case where $R_{\text{sec}}$ prevents a risky shortcut. These examples make the mechanism legible and connect the quantitative metrics back to concrete programmer-visible behavior.

\subsection{Limitations}

There are several limitations to keep in mind:
\begin{itemize}
    \item \textbf{Small evaluation set.} The reported 5\% test success comes from 1/20 prompts in this pilot.
    \item \textbf{Security findings rarity (in pilot eval).} Bandit findings were absent in the evaluated samples, so the security component did not meaningfully differentiate models in this run.
    \item \textbf{Competitive-programming bias.} APPS+ emphasizes stdin/stdout and algorithmic correctness; results may not transfer directly to other coding tasks (APIs, libraries, applications).
\end{itemize}

\section{Conclusion}

This paper presents SecureCodeRL, a security-aware RL pipeline for code generation with a combined objective $R=\alpha R_{\text{func}}+\beta R_{\text{sec}}$ and a partial-credit functional reward to mitigate sparse test feedback. I first motivate the problem with a multi-model benchmark (Table~\ref{tab:benchmark}) and a compact failure taxonomy (Table~\ref{tab:failtax}) that explains why binary rewards are brittle. In a small pilot evaluation on APPS+, PPO with partial credit (using a continued-training variant) achieves the strongest results among the variants I tested: higher syntax validity and the only non-zero test success signal in this pilot, while remaining clean under Bandit static analysis. The main takeaway is not that the problem is solved; it is that reward shaping can meaningfully change whether PPO learns anything at all in strict unit-test environments. The next step is scaling evaluation and adding ablations to turn this proof-of-concept into a fully benchmarked result.

\bibliographystyle{IEEEtran}

\end{document}